\documentclass[aps,prb,showpacs,twocolumn]{revtex4}

\usepackage{graphicx}
\usepackage{amsmath}
\usepackage{natbib}
\usepackage{bm}
\usepackage[dvips]{color}

\begin{document}

\title{Dynamic Lattice Distortions in Sr$_2$RuO$_4$: A microscopic study by perturbed angular correlation (TDPAC) spectroscopy}
\date{\today}

\author{S.N.~Mishra}
\email[Electronic address: ]{mishra@tifr.res.in}
\affiliation{Department of Nuclear and Atomic Physics, Tata Institute of Fundamental Research, Homi Bhabha Road,
Mumbai-400005, India}

\author{M.~Rots}
\affiliation{Instituut voor Kern- en Stralingsfysica and INPAC, Katholieke Universiteit Leuven, Celestijnenlaan 200 D, BE-3001
Leuven, Belgium}

\author{S.~Cottenier}
\email[Electronic address: ]{Stefaan.Cottenier@fys.kuleuven.be}
\affiliation{Instituut voor Kern- en Stralingsfysica and INPAC, Katholieke Universiteit Leuven, Celestijnenlaan 200 D, BE-3001
Leuven, Belgium}

\begin{abstract}

Applying time differential perturbed angular correlation (TDPAC) spectroscopy and \emph{ab~initio} calculations, we have
investigated possible lattice instabilities in Sr$_{2}$RuO$_{4}$ by studying the electric quadrupole interaction of a $^{111}$Cd
probe at the Ru site. We find evidence for a dynamic lattice distortion, revealed from the observations of: (i) a rapidly
fluctuating electric-field gradient (EFG) tensor of which the main component decreases with decreasing temperature, and (ii) a
monotonic increase of the EFG asymmetry ($\eta$) below 300~K. We argue that the observed dynamic lattice distortion is caused by
strong spin fluctuations associated with the inherent magnetic instability of Sr$_{2}$RuO$_{4}$.

\end{abstract}

 \maketitle

\section{Introduction}

Since the discovery of unconventional superconductivity in the non-cuprate oxide Sr$_{2}$RuO$_{4}$, considerable attention has
been paid to study its electronic and magnetic properties in the normal and superconducting states.\cite{Maeno1, Andrew} This
material -- with a layered perovskite structure identical to the prototype high-T$_{c}$ superconductor (La, Sr/Ba)$_{2}$CuO$_{4}$
-- has been identified to exhibit spin triplet (p-wave) superconductivity below 1~K.\cite{Maeno1, Andrew} Besides
superconductivity, Sr$_{2}$RuO$_{4}$ exhibits several other unusual features in its normal state. For instance, transport
studies have revealed Fermi liquid behavior with an electrical resistivity that shows quadratic temperature dependence below
25~K.\cite{Maeno2} Secondly, the material shows a large electronic specific heat ($\gamma \approx$ 37.5 mJ/mol-K$^{2}$ and a strongly enhanced Pauli susceptibility ($\chi_{0} \approx 9\times 10^{-4}$emu/mol) reflecting heavy-fermion behavior.\cite{Maeno2, Nakatsuji} It has been recognized that both structural and magnetic instabilities play a crucial role for the unconventional superconductivity in Sr$_{2}$RuO$_{4}$.\cite{Maeno1, Imai} Recent theoretical calculations have suggested that Sr$_{2}$RuO$_{4}$ is close to a ferromagnetic (FM) instability\cite{Rice, Fang} with strong FM spin fluctuations which may lead to spin triplet p-wave superconductivity.\cite{Mazin}

As for the structural aspect, while crystal transformation is common in most cuprate oxides, no long range structural
distortion/transformation has been detected for Sr$_{2}$RuO$_{4}$ down to 100~mK.\cite{Gardner, Vogt, Neumeier} However, a local
(short-ranged) structural instability related to a rotation of the RuO$_{6}$ octahedra has been detected by Braden et al., using
phonon dispersion information obtained by inelastic neutron scattering.\cite{Braden} To our knowledge, no other experimental
study related to structural instabilities in Sr$_{2}$RuO$_{4}$ has been reported. Considering the many unusual physical properties, including unconventional superconductivity, it is desirable to carry out additional investigations, particularly with help of microscopic techniques, to examine structural distortions/instabilities in undoped Sr$_{2}$RuO$_{4}$. For this
purpose, hyperfine methods such as M\"{o}ssbauer spectroscopy, Nuclear Magnetic/Quadrupole Resonance (NMR/NQR) or
Time-Differential Perturbed Angular Correlation (TDPAC) are extremely useful. In particular, information on structural
properties such as lattice transformation and/or distortion can be extracted by studying the electric-field gradient (EFG) tensor
obtained from nuclear quadrupole interaction measurements. Performing such studies with NMR or NQR has been found to be
difficult, due to a.o.\ low isotopic abundance, small quadrupole moments, strong spin lattice relaxation and large line broadening. \cite{Mukuda, Ishida-Minami, Ishida-Kitaoka} Similarly, due to the large line width, M\"{o}ssbauer spectroscopy using the native $^{99}$Ru isotope is also not suitable for measuring small lattice distortions. Such experimental difficulties, however, can be overcome using the TDPAC method\cite{SchWe}. With a suitable probe nucleus such as $^{111}$Cd, the TDPAC technique can be very effective for detection of static and dynamic lattice distortions.

In this paper we report our results on a lattice instability in Sr$_{2}$RuO$_{4}$ observed from quadrupole interaction
measurements at a $^{111}$Cd probe nucleus measured by TDPAC spectroscopy. The experimental results show evidence of a
\emph{dynamic} lattice distortion below 300~K revealed by: (i) a rapidly fluctuating electric-field gradient (EFG) tensor and (ii) a temperature dependent change of the EFG asymmetry parameter $\eta$. Taking into account \emph{ab~initio} calculations based on Density Functional Theory (DFT), we argue that the observed dynamic lattice distortion is caused by strong spin fluctuations associated with an inherent magnetic instability of Sr$_{2}$RuO$_{4}$.

\section{Experimental and Computational Details}

A polycrystalline sample of Sr$_{2}$RuO$_{4}$ was prepared following a standard procedure. A mixture of stoichoimetric
amounts of high purity SrCO$_{3}$ and RuO$_{2}$ was pelletized and sintered at 1100$^\circ$C for 12~h. After regrinding, the powder was pressed into a pellet and heated in air at 1100$^\circ$C for another 18~h followed by furnace cooling. The sample was
characterized by powder X-ray diffraction, showing a single phase having the K$_{2}$NiF$_{4}$-type tetragonal structure with lattice papameters a = 3.868(5)~\AA\ and c = 12.732(5)~\AA\, which agree well with the earlier data.\cite{Maeno2, Gardner, Vogt, Neumeier} The $^{111}$Cd probe was introduced into a small piece of Sr$_{2}$RuO$_{4}$ by diffusing carrier free parent $^{111}$In
activity at 1050$^\circ$C for 24~h followed by rapid cooling to room temperature. It is important to mention that the typical
concentration of Cd in the sample remains below 1~ppm and thus does not alter the basic physical properties of the material. The
electric quadrupole interaction parameters were extracted from the life time spectra of the 247 keV, I=5/2 and 84~ns level of the
daughter $^{111}$Cd nucleus produced by electron capture (EC) decay of $^{111}$In. The time spectra were recorded simultaneously
in a 90$^\circ$/180$^\circ$ geometry using a set-up consisting of four BaF$_{2}$ detectors and a standard slow-fast coincidence
circuit having a time resolution better than 700 ps. Measurements were carried out as a function of temperature in the range 20-500 K using either a closed cycle helium refrigerator or a specially designed furnace. The perturbation factor A$_{22}$G$_{22}(t)$ in the angular correlation function $W(\theta, t) = \sum_{kk}A_{kk}G_{kk}(t)P_{k}(cos\theta)$ was obtained by
constructing the appropriate ratio function, called the \mbox{TDPAC} time spectrum \cite{SchWe},
\begin{eqnarray*}
{R(t) \ = \ A_{22}G_{22}(t) \ = \frac
{2[W(180^\circ,t)-W(90^\circ,t)]}{[W(180^\circ,t)+2W(90^\circ,t)]}}
\end{eqnarray*}
\noindent where $W(\theta,t)$ are the background-subtracted normalized coincidence counts of detectors placed at 180$^\circ$
and 90$^\circ$. The spectra were fitted to the function \cite{SchWe}

\begin{eqnarray*}
{A_{22}G_{22}(t) =
\frac{A_{22}}{5}[1+\sum_{n=1}^{3}S_{n}(\eta)e^{-nt/\tau_N}\cos(n\omega_{0}(\eta)t)]}
\end{eqnarray*}
to extract the interaction frequency $\omega_{0}$ related to the principal component V$_{\mathrm{zz}}$ of the EFG
tensor\cite{SchWe}:
\begin{eqnarray}
\label{eq-omega0} \omega_0 & = & \frac{6eQV_{zz}}{\hbar \, 4I(2I-1)}
\end{eqnarray}
(expression valid for half-integer spin), the asymmetry $\eta=V_{\mathrm{xx}}-V_{\mathrm{yy}}/V_{\mathrm{zz}}$ of the EFG
tensor, and the relaxation time $\tau_N$ which is related to the distribution width $\delta$ of $\omega_0$ by 
$\tau_N = 1/\delta$. For our data, using a Lorenzian frequency distribution yielded a better $\chi^{2}$ than a Gaussian spread in $\omega_0$.

Complimentary to the TDPAC experiments, we performed a series of \emph{ab initio} calculations on pure and Cd-doped Sr$_2$RuO$_4$. The calculations were performed within Density Functional theory\cite{Hohenberg1964,Sham1965,DFT-LAPW2002}, using the Augmented Plane Waves + local orbitals (APW+lo) method\cite{Sjostedt2000,Madsen2001,DFT-LAPW2002} as implemented
in the WIEN2k package\cite{Wien2k} to solve the scalar-relativistic Kohn-Sham equations. In the APW+lo method, the
wave functions are expanded in spherical harmonics inside nonoverlapping atomic spheres of radius $R_{\mathrm{MT}}$, and in
plane waves in the remaining space of the unit cell (=the interstitial region). For Sr, Ru and Cd a $R_{\mathrm{MT}}$ value
of 1.85 a.u.\ was chosen, for O we used $R_{\mathrm{MT}}=$1.55 a.u. The maximum $\ell$ for the expansion of the wave function in
spherical harmonics inside the spheres was taken to be $\ell_{\mathrm{max}}=10$. The plane wave expansion of the wave
function in the interstitial region was made up to $K_{\mathrm{max}}=7.5/R_{\mathrm{MT}}^{\mathrm{min}}=4.83$~a.u.$^{-1}$
for pure Sr$_{2}$RuO$_{4}$ and for supercells with 8 formula units, and up to $K_{\mathrm{max}}=5.75/R_{\mathrm{MT}}^{\mathrm{min}}=3.71$~a.u.$^{-1}$ for supercells with 16 formula units (this reduced accuracy has an effect on e.g.\ the electric-field gradient of less than 10\%). The charge density was Fourier expanded up to $G_{\mathrm{max}}=16 \sqrt{Ry}$. For the sampling of the Brillouin zone, a special k-mesh equivalent to 8$\times$8$\times$8 mesh in the pure Sr$_{2}$RuO$_{4}$ structure was used throughout. As exchange-correlation functional, the Perdew-Burke-Ernzerhof
Generalized Gradient Approximation (GGA) was used.\cite{GGA} As Table~\ref{tab2} shows, this type of calculations is able to
reproduce correctly some experimentally known structural and EFG properties of pure Sr$_2$RuO$_4$. The structural information is
also in agreement with previous \emph{ab~initio} calculations,\cite{DJS} while for the EFG's our GGA results show
much better agreement with respect to experiment than reported values obtained using the Local Density Approximation.\cite{KBHH}
\begin{table}
 \begin{center}
  \caption{Comparison of some structural and electric-field gradient parameters for I4/mmm Sr$_2$RuO$_4$ with fully optimized theoretical lattice constants and internal positions (left), and the corresponding experimental data (right).
\label{tab2}}
  \begin{ruledtabular}
   \begin{tabular}{l|ll}
    &   theory & experiment \\ \hline
    a=b (\AA) & 3.8937 & 3.862 [\onlinecite{Vogt}] \\
    c (\AA) & 12.8935 & 12.722 [\onlinecite{Vogt}] \\
z$_{\mathrm{Sr}}$ & 0.3525 & 0.3534 [\onlinecite{Vogt}] \\
z$_{\mathrm{O(2)}}$ & 0.1626 & 0.1613 [\onlinecite{Vogt}] \\
d$_{\mathrm{Sr}-\mathrm{Ru}}$ (\AA) & 3.3464 & 3.3069 [\onlinecite{Vogt}] \\
d$_{\mathrm{Ru}-\mathrm{O(2)}}$ (\AA) & 2.0962 & 2.0521 [\onlinecite{Vogt}] \\
V$_{\mathrm{zz}}$ Sr (10$^{21}$ V/m$^2$) & -2.42 & --       \\
V$_{\mathrm{zz}}$ Ru (10$^{21}$ V/m$^2$) & -0.99 & $\pm$2.05[\onlinecite{Ishida-Kitaoka}] \\
V$_{\mathrm{zz}}$ O(1) (10$^{21}$ V/m$^2$) & +7.90 & $\pm$8.1 [\onlinecite{Mukuda}]\\
V$_{\mathrm{zz}}$ O(2) (10$^{21}$ V/m$^2$) & +7.10 & $\pm$6.4 [\onlinecite{Mukuda}]\\
$\eta$ O(1)  & 0.265 & 0.17 [\onlinecite{Mukuda}]\\
   \end{tabular}
  \end{ruledtabular}
 \end{center}
\end{table}

\section{Results and Discussion}

\begin{figure}
\includegraphics{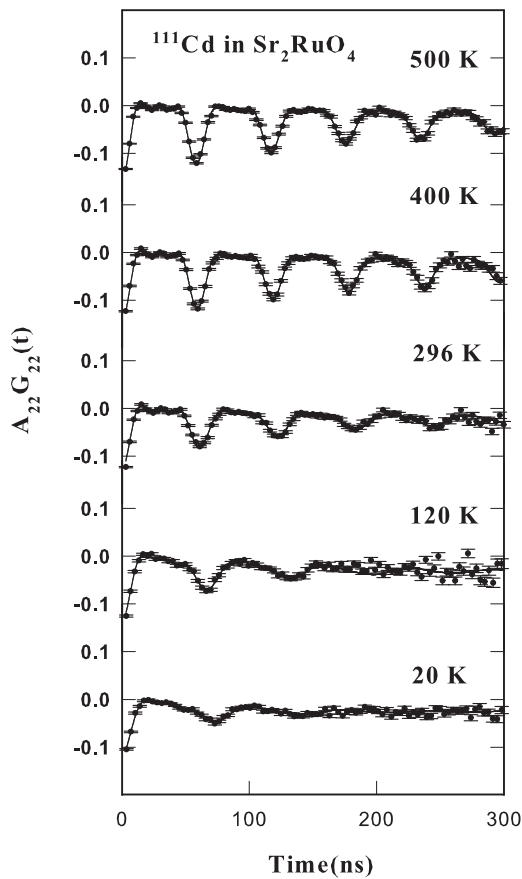}
\caption{Typical TDPAC spectra for $^{111}$Cd in Sr$_{2}$RuO$_{4}$ at different temperatures.} \label{1}
\end{figure}

\begin{figure}
\includegraphics{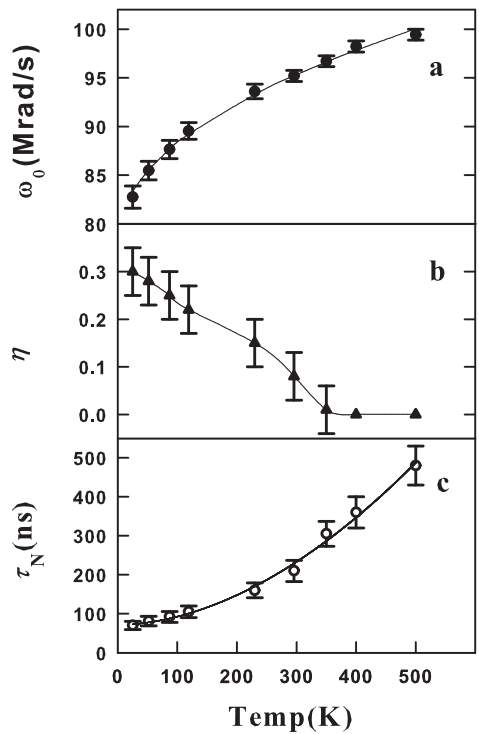}
\caption{Temperature dependence of the quadrupole interaction frequency ($\omega_{0}$), electric field gradient asymmetry parameter ($\eta$), and the nuclear relaxation time ($\tau_{N}$) measured for $^{111}$Cd in Sr$_2$RuO$_4$. The solid line in (b) is a guide to the eye whereas in (a) and (c) they correspond to least square fits to equations discussed in the text.} \label{2}
\end{figure}

Figure~\ref{1} shows some typical TDPAC spectra of $^{111}$Cd in Sr$_2$RuO$_4$ at different temperatures.  All spectra show a
single quadrupole interaction frequency with nearly the full anisotropy (A$_{22}~\approx~$0.13), indicating that the $^{111}$Cd
probe atoms occupy a unique lattice site. From a comparison of the chemical behaviour and ionic size of the mother isotope $^{111}$In and the atomic species of the sample under investigation, the Cd probe atoms are likely to occupy a substitutional Ru site in Sr$_2$RuO$_4$. This assignment of lattice site is supported by earlier TDPAC results in related materials, e.g.\ SrRuO$_{3}$ and CaRuO$_{3}$\cite{Catchen} where Cd has been reported to appear at the Ru site, and is corroborated by the EFG-values found from our \emph{ab~initio} calculations (see Tab.~\ref{tab3} and the corresponding discussion).
The spectra recorded above 300~K could be simulated with a single randomly oriented, axially symmetric ($\eta$=0) electric-field
gradient (EFG). From Eq.~\ref{eq-omega0} and using the experimental quadrupole moment Q = 0.83~b\cite{Pramila}, the
principal component of the EFG at room temperature was determined to be V$_{zz} = 5.2(1)\times 10^{21}V/m^{2}$. Below 300~K the
\mbox{TDPAC} spectra show a noticeable deviation from the axially symmetric quadrupole interaction patterns observed at higher
temperatures (Fig.~\ref{1}). They could be fitted with a randomly oriented, asymmetric EFG yielding $\omega_{0}$ = 82.7(11) Mrad/s and asymmetry parameter $\eta$=0.30(3) at 20~K. In addition, the observed R(t) spectra show temperature dependent damping, most pronounced at low temperatures (see Fig.~\ref{1}). Fig.~\ref{2} displays the variation of $\omega_{0}$, $\eta$ and the relaxation time $\tau_{N}=1/\delta(\omega_{0})$ as a function of temperature (see also Table~\ref{tab1}). The relaxation time $\tau_N$ measures the spread in interaction frequencies, and therefore is an indication of the number of different environments felt by the probe nucleus.
\begin{table}
\begin{center}
\caption{Summary of quadrupole interaction parameters for $^{111}$Cd in Sr$_2$RuO$_4$. \label{tab1}}
\begin{tabular}{|c|c|c|c|c|}
\hline Temperature & $\omega_{0}$ &  V$_{zz}$ & $\eta$ & $\tau_{N}$ \\

(K)& (Mrad/s) & ($10^{21}V/m^{2}$) &  & (ns) \\
\hline
20 & 82.7(9) & 4.56(10) & 0.30(5) & 70(20) \\
52 & 85.5(8) & 4.71(10) & 0.27(5) & 81(20) \\
87 & 87.7(7) & 4.84(10) & 0.25(5) & 95(20) \\
119& 90.0(5) & 4.97(10) & 0.22(5) & 115(20)\\
230& 94.0(3) & 5.03(10) & 0.15(5) & 170(25)\\
296& 95.2(3) & 5.20(10) & 0.08(5) & 220(35)\\
350& 96.7(2) & 5.11(10) & 0.0     & 310(40)\\
400& 98.2(2) & 5.19(10) & 0.0     & 380(50)\\
500& 99.5(2) & 5.26(10) & 0.0     & 480(60)\\
\hline
\end{tabular}
\end{center}
\end{table}

The TDPAC results obtained for $^{111}$Cd in Sr$_2$RuO$_4$ (Fig.~\ref{2} and Table~\ref{tab1}) reveal several interesting
features. First of all, the quadrupole interaction frequency $\omega_{0}$ \emph{increases} from 82.7(11)~Mrad/s at 20~K to
almost 100~Mrad/s at 500~K. This variation of $\omega_{0}$ as a function of temperature could be parameterized using the relation
\begin{eqnarray*}
{\omega_{0}(T) = \omega_{0}(0)(1 + A T^{\alpha})}
\end{eqnarray*}
with $\omega_{0}(T=0)$ = 77.8(14)~Mrad/s (or V$_{\mathrm{zz}}(T=0)=4.2\times10^{21}~$V/m$^2$), A = 0.016(5) and
$\alpha$ = 0.47(6). Considering that Sr$_2$RuO$_4$ is a metallic system\cite{Maeno1, Andrew}, the quadrupole interaction frequency is expected to \emph{decrease} with increasing temperature, following a T$^{3/2}$ law due to lattice
vibrations.\cite{Kaufmann, Torumba} Instead, the $\omega_{0}$ measured for $^{111}$Cd in Sr$_{2}$RuO$_{4}$ \emph{increases} with
temperature and approximately shows a $\sqrt T$ dependence.

Secondly, while the EFG is observed to be axially symmetric above 300 K ($\eta=0$), it shows a significant amount of asymmetry at
low temperatures with $\eta(T)$ monotonically increasing up to $\approx$0.3 at 20~K. These observed $\eta(T)$ data provide a
direct proof that i) the local site symmetry of the Cd probe atom is \emph{not} cubic as expected for the Ru site in this lattice
structure, and ii) continuously changes below 300~K. 

Thirdly, the relaxation time $\tau_{N}$ extracted from the damping of the R(t) spectra strongly varies with temperature (Fig.~\ref{1} and Fig.~\ref{2}c). Such a spread in interaction frequencies can be due either to a time dependent fluctuation of the EFG and/or to a static distribution of EFG values due to a spread in possible environments of the Cd probe. Although it is difficult to rule out the latter, in view of the facts that the interaction frequencies show a Lorentz distribution and the relaxation time is strongly temperature dependent (which would be hard to imagine for a static distribution), we conclude that the observed damping is caused by a dynamic fluctuation of the EFG at the Cd site. A large $\tau_N$ at high temperatures means that all probe nuclei feel quite similar environments: the EFG fluctuations are fast enough to nearly average out over the life time of probe. At low temperature, the spread in interaction frequencies appears to be much larger: slower EFG fluctuations cause that every probe nucleus becomes sensitive to the details of the local fluctuation history. The temperature variation of $\tau_N$, summarized in Table~\ref{tab1}, could be fitted by the relation
\begin{eqnarray*}
{\tau_{N}(T) = \tau_{N}(0)(1 + B T^{\beta})}
\end{eqnarray*}
with $\tau_{N}(0) = 71$ ns and $\beta = 1.85$, revealing a nearly quadratic dependence.

How can we explain these three experimental observations? A straightforward suggestion to explain a fluctuating EFG with
non-zero $\eta$ would be to assume migration of Cd to other sites of lower symmetry and/or hopping/diffusion of oxygen from the
CdO$_6$ octahedron. The former, however, is unlikely because in Sr$_2$RuO$_4$ the only substitutional site without axial symmetry
is O(1) and it is highly improbable that Cd will replace an O-atom when chemically similar atoms as Sr and Ru are present. For the same reason, the occupation of an interstitial site is unlikely. Furthermore, in the case of site migration (diffusive/hopping motion) it is generally observed that the $\tau_{N}$(T) shows an activation type (Arrhenius) behaviour\cite{Karlsson} in contrast to the power law dependence observed in the present case. Another way to explain the appearance of a non-zero $\eta$ would be to invoke a long-range (e.g.\ Jahn-Teller like) lattice distortion.
This possibility is excluded, however, by structural studies using high-resolution X-ray as well as neutron diffraction
techniques.\cite{Gardner, Vogt, Neumeier} Both have failed to detect any lattice transformation or distortion in Sr$_2$RuO$_4$
down to 100~mK. As a result, only \emph{short ranged} lattice distortions are allowed. In order to account for the fluctuations
we have observed from the EFG, these distortions must also be \emph{dynamic}. Our TDPAC results thus provide clear signatures of
a dynamic lattice distortion in Sr$_2$RuO$_4$ below 300~K. The dynamic lattice distortion observed from our TDPAC measurements
are consistent with the structural instability reported from phonon dispersion results observed by inelastic neutron
scattering,\cite{Braden} although there are subtle differences with respect to the temperature dependence. For instance, we
observe that above 300~K these symmetry-breaking lattice distortions are varying fast enough to be averaged out over the
time-window of our experiment (100-200~ns).

What could be the physical origin of the dynamic lattice distortions in Sr$_2$RuO$_4$, observed through our TDPAC
experiments? It is now well established that the highly correlated material Sr$_2$RuO$_4$ lies close to a magnetic
instability.\cite{Mazin, Fang} Recently, Fang and Terakura\cite{Fang} showed by first principle calculations based
on the local spin density approximation (LDA) that the type of magnetic structure of Sr$_2$RuO$_4$ strongly influences its
lattice structure. More precisely, the magnetic order influences the details of the RO$_6$ octahedrons (their tilt ($\theta$),
their rotation ($\phi$) and their flatness $\lambda$, the latter defined as the ratio of the Ru-O bond length along the c-direction and along the a- or b-direction ($\lambda = d_{c}/d_{ab})$). For the experimentally observed stable structure of Sr$_2$RuO$_4$ the ground state was found to be ferromagnetic, but non-magnetic and antiferromagnetic solutions with very small differences in total energy were found as well. Their calculated results revealed that Sr$_2$RuO$_4$ can be easily driven into different magnetic states by a small distortion of the lattice, especially by modifying the RuO$_{6}$ octahedron. We apply this argument in the opposite way: if -- for instance as a result of spin fluctuations -- the local magnetic structure of Sr$_2$RuO$_4$ will change continuously, it will be accompanied by a local and continuously varying distortion. This gives rise both to the fluctuating EFG and the deviation from cubic symmetry. As these dynamic distortions are mediated by magnetic effects (spin fluctuations), they differ from the usual lattice dynamics. As a consequence, it is not surprising that we observe an anomalous temperature dependence of the EFG. At the lowest studied temperatures, the spin fluctuations are slow
enough to provide every Cd probe with different sequence of fluctuations during its life time, leading to a large spread in
observed interaction frequencies. Near room temperature, the spin fluctuations (and therefore the dynamic lattice distortions as
well) are so fast that they average out over the relevant time window, and an averaged, axially symmetric environment is
detected. Such short-ranged and time dependent lattice distortions can not be detected through X-ray or neutron diffraction
experiments, which measure the long range correlation of atomic arrangements in solids, averaged over large time intervals and
therefore insensitive to dynamic changes over short length scales. On the other hand, as demonstrated here, the affects of short
range dynamic lattice modifications can manifest themselves in experiments by microscopic techniques such as TDPAC.

Further support for our interpretation of short-ranged dynamic lattice changes triggered by spin fluctuations comes from a series
of \emph{ab~initio} calculations. We calculated the main component V$_{\mathrm{zz}}$ of the electric field gradient tensor for Cd in Sr$_2$RuO$_4$, using a super cell to mimic the condition of an isolated impurity. The presence of a Cd impurity influences the positions of its neighboring Sr, Ru and O atoms, and we allowed the first three shells of Cd-neighbours (twice O and once Ru)  to move to their new equilibrium positions. More distant neighbours hardly moved. Tests with different sizes of supercells showed that it was necessary to take a cell with 1 Cd atom per 16 formula units of Sr$_2$RuO$_4$ ($2\sqrt{2}\times2\sqrt{2}\times1$ supercell, containing 112 atoms). In such a supercell, the Cd atoms occupy a tetragonal (but nearly cubic) sublattice with Cd-Cd separations of 11.0~\AA\ and 12.9~\AA\, which is reassuringly large so that impurity-impurity interactions can be neglected \cite{Cottenier-2004}. The results obtained from these calculations, both for the unrelaxed (all atoms at ideal I4/mmm lattice sites) and the relaxed conditions are summarized in Table~\ref{tab3}. It must be mentioned that the lattice relaxations used in our calculations are only approximate as it was performed in a supercell with 1 Cd atom per 8 formula units, and the distances were carried over to the larger supercell. Calculating relaxations directly in the largest supercell was computationally too expensive.
\begin{table}
 \begin{center}
  \caption{V$_{\mathrm{zz}}$ (10$^{21}$ V/m$^2$) for a
  Cd impurity at the Ru-position in unrelaxed and relaxed supercells with
  16 formula units of Sr$_2$RuO$_4$ per Cd, for non-magnetic and antiferromagnetic calculations.
  The range of experimental values cover the interval 0~K -- 500~K (see Fig~\ref{2}).
  \label{tab3}}
  \begin{ruledtabular}
   \begin{tabular}{l||l|l|l}
     &   non-magnetic & \multicolumn{1}{c}{AF} & \multicolumn{1}{c}{FM} \\ \hline
    V$_{\mathrm{zz}}$ / $\eta$ Cd (unrelaxed) & 2.2/0.00 & 1.9/0.09 & 1.5/0.00 \\
    V$_{\mathrm{zz}}$ / $\eta$ Cd (relaxed) & 1.1/0.00 & 4.2/0.01 & 4.5/0.00  \\
    V$_{\mathrm{zz}}$ / $\eta$ Cd (exp) & \multicolumn{3}{c}{4.3 -- 5.5 / 0.0 -- 0.3}
    \\
   \end{tabular}
  \end{ruledtabular}
 \end{center}
\end{table}

If no spin-polarization no r lattice relaxation is allowed, the main component of the EFG tensor at a Cd impurity substituting Ru
is calculated to be V$_{\mathrm{zz}}=2.2~10^{21}~$V/m$^2$ (Tab.~\ref{tab3}). Including lattice relaxation, the EFG decreases
to V$_{\mathrm{zz}}=1.1~10^{21}~$V/m$^2$. Both these values are smaller compared to the range of experimental values observed in
the temperature interval 20--500~K. The EFG tensor is in both cases axially symmetric ($\eta=0$), due to the point group
symmetry at the Ru site. For the spin-polarized calculations, we tried two different spin configurations: antiferromagnetic (AF)
and ferromagnetic (FM). Without lattice relaxations, the value of V$_{\mathrm{zz}}$ gets somewhat smaller compared to the
non-magnetic case. It becomes considerably larger, however, if lattice relaxations are allowed: the magnetism does not affect the
EFG directly, but it affects the lattice relaxation, which in turn affects the EFG. Most importantly, the order of magnitude
(4.2-4.5) of V$_{\mathrm{zz}}$ is now perfectly comparable to the experimental observation (4.2 when extrapolated to 0~K). This
corroborates our assumption that Cd indeed substitutes a Ru atom. If the applied spin configuration lacks axial symmetry (as is the case for our AF configuration, and as will be the case for any `random' spin configuration that is the momentaneous result of
fluctuating spins), the calculations show a non-zero $\eta$ does indeed appear. The calculated values of 0.09 and 0.01 are smaller than the observed maximum of 0.30, but these values are for one particular AF configuration only. Taken all together, these \emph{ab~initio} calculations demonstrate that different spin configuration (as they appear during the spin fluctuation process) do indeed influence the EFG tensor at the Cd site, and lead to better numerical agreement between the calculated and measured EFG tensor.

In summary, by studying the quadrupole interaction of $^{111}$Cd using the TDPAC technique and \emph{ab~initio} calculations, we
have found evidence for a dynamic lattice distortion in Sr$_2$RuO$_4$ reflected by a rapidly fluctuating electric field
gradient (EFG) and a temperature dependent change of its asymmetry parameter ($\eta$). The results presented in this work might
contribute to understanding the mechanism of superconductivity in Sr$_2$RuO$_4$.

\acknowledgments

Part of this work has been financially supported by Project No. G.0237.05 of the Fonds voor Wetenschappelijk Onderzoek --
Vlaanderen (FWO), the Concerted Action Programme of the K.U.Leuven (GOA/2004/02), the Centers of Excellence Programme of the
K.U.Leuven (INPAC, EF/05/005) and the Inter-University Attraction Pole Programme (IUAP P5/1).

\end{document}